# CALCULATIONS OF HADRONIC MATRIX ELEMENTS USING LATTICE QCD

Rajan Gupta

T-8, MS-B285, Los Alamos National Laboratory, Los Alamos, NM 87545

I give a brief introduction to the scope of lattice QCD calculations in our effort to extract the fundamental parameters of the standard model. This goal is illustrated by two examples. First I discuss the extraction of CKM matrix elements from measurements of form factors for semileptonic decays of heavy-light pseudoscalar mesons such as $D \to Ke\nu$. Second, I present the status of results for the kaon $B$ parameter relevant to CP violation. I conclude the talk with a short outline of our experiences with optimizing QCD codes on the CM5.

## 1. Introduction

Current high energy experiments show that the fundamental building blocks of matter are quarks, gluons, leptons, photons, weak bosons and the elusive Higgs particle. The interactions between these particles are described by a set of theories, known collectively as the Standard Model. While this model has been immensely successful, and present data do not demand enhancements to the model or a new theory altogether, it is still incomplete. Experimentalists have yet to discover the top quark, the $\tau$ neutrino and the Higgs boson. On the other hand it has proven very difficult to extract the predictions of the Standard Model when the interactions among the elementary particles are strong. This happens in processes in which quarks interact through the exchange of gluons carrying 4-momenta less than a few GeV. Such processes cannot be calculated reliably using perturbation theory as there is no small expansion parameter. For this reason it has proven extremely difficult to make precise quantitative tests of the theory, such as making quantitative predictions that can be compared to experiments. Even twenty years after the formulation of QCD as the theory of strong interactions this state of affairs persists. What one needs are non-perturbative tools to include strong interaction effects. At present the most promising approach is to carry out large-scale numerical simulations using a lattice version of the gauge theory. In this talk I hope to describe the computational challenge presented by lattice QCD and the progress we have made.

Let me begin by enumerating the 24 parameters of the standard model.

| Parameters | Number | Comments |
|---|---|---|
| Masses of quarks | 6 | $u$, $d$, $s$ light |
| | | $c$, $b$ heavy |
| | | $t > 90\ GeV$ ?? |
| Masses of leptons | 6 | $M_{e,\ \mu,\ \tau}$ known |
| | | $M_{\nu_e,\ \nu_\mu,\ \nu_\tau} = 0$ ?? |
| Mass of $W^\pm$ | 1 | 81 GeV |
| Mass of $Z$ | 1 | 92 GeV |
| Mass of gluons, $\gamma$ | 1 | 0 |
| Mass of Higgs | 1 | Not Found |
| Coupling $\alpha_s$ | 1 | $\approx 1$ for Energy $< 1$ GeV |
| Coupling $\alpha_{em}$ | 1 | 1/137 |
| Coupling $G_F = \frac{\sqrt{2}g_w^2}{8M_W^2}$ | 1 | $10^{-5}\ Gev^{-2}$ |
| Weak Mixing Angles | 3 | $\theta_{12}, \theta_{23}, \theta_{13}$ |
| CP Violating phase | 1 | $\delta$ |
| Strong CP parameter | 1 | $\Theta = 0$ ?? |

Of these parameters the ones whose determination requires input from lattice QCD are the masses of light quarks, $m_u$, $m_d$, $m_s$, the strong coupling $\alpha_s$, the weak mixing angles and the CP violating phase $\delta$, and the strong CP parameter $\Theta$. Precise determination of their values will either validate the standard model or provide clues to new physics.

The weak mixing angles and the CP violating phase $\delta$ need some introduction. These parameters arise because quarks are not eigenstates of weak-interactions. The mixing between flavors is described by the $3 \times 3$ Cabibbo-Kobayashi-Maskawa ($CKM$) matrix $V$,

$$V = \begin{pmatrix} V_{ud} & V_{us} & V_{ub} \\ V_{cd} & V_{cs} & V_{cb} \\ V_{td} & V_{ts} & V_{tb} \end{pmatrix} \ .$$

Here, for example, $V_{ub}$ is the strength of $b \to u$ flavor transformation as a result of charged $W$ exchange. For 3 generations $V^{-1} = V^\dagger$ and the matrix can be written in terms of 4 independent parameters, the 3 angles $\theta_{12}, \theta_{23}$ and $\theta_{13}$ and the CP violating phase $\delta$ as [1]

$$V = \begin{pmatrix} c_{12}c_{13} & s_{12}c_{13} & s_{13}e^{-i\delta} \\ -s_{12}c_{23} - c_{12}s_{23}s_{13}e^{i\delta} & c_{12}c_{23} - s_{12}s_{23}s_{13}e^{i\delta} & s_{23}c_{13} \\ s_{12}s_{23} - c_{12}c_{23}s_{13}e^{i\delta} & -c_{12}s_{23} - s_{12}c_{23}s_{13}e^{i\delta} & c_{23}c_{13} \end{pmatrix}$$

where $c_{ij} = \cos\theta_{ij}$ and $s_{ij} = \sin\theta_{ij}$ for $i = 1, 2, 3$. A non-zero value of $\delta$ gives rise to CP violation in weak decays.

The strong CP violating parameter $\Theta$ arises because there is no symmetry or dynamical argument to rule out a term like $\mathcal{L}_\Theta = (i\Theta g^2/32\pi^2)F\widetilde{F}$ from the QCD Lagrangian. Even though this term is a total divergence its presence leads to observable consequences like CP violation because of instanton solutions in QCD. The best bound on this parameter $\Theta < 10^{-9}$ comes from measurements of the electric dipole moment of the neutron, $d_N < 1.2 \times 10^{-25} e$ cm [2].

The crucial matrix element needed in the theoretical analysis is of the pseudoscalar density $\overline{u}\gamma_5 u + \overline{d}\gamma_5 d + \overline{s}\gamma_5 s$ within the neutron, and lattice calculations hope to provide a non-perturbative estimate. At present the numerical technology is not sufficiently well developed to undertake this calculation; what needs to be done is described in Ref. [3] and I refer to it for details.

To set the stage for the results presented later, let me give an outline of how lattice QCD interfaces with experimental data and theoretical predictions of the standard model to test the theory. The general form of SM prediction for a process is an expression (which I will call the master equation) consisting of three parts; known factors times some function of the unknown parameters times the matrix element of the appropriate operator sandwiched between initial and final states. Thus for each process for which there exists accurate experimental data, knowing the value of the matrix element gives an equation of constraint for the remaining part involving the unknown parameters. Once a certain number of such calculations are in hand we can extract accurate values for all the unknown parameters. Thereafter the standard model can be used to make accurate predictions for other processes. In this talk I will demonstrate this strategy with two examples, semi-leptonic form-factors and the kaon $B$ parameter, that are discussed in Sections 6 and 7 respectively.

I will assume that the reader is familiar with Monte Carlo methods and Lattice QCD. Those who are not should, at this point, read the excellent pedagogical introduction given by D. Toussaint at this meeting or the monogram by Creutz [4].

## 2. Errors in lattice calculations

Lattice calculations rely on a Monte Carlo sampling of configurations generated on a discrete space-time grid. Correlation functions are calculated as a statistical average, and are composed of gauge variables defined on links and quark propagators calculated on these background gauge configurations. This procedure introduces statistical and systematic errors into the results, so in order for you to judge progress in the field it is important for me to first explain these sources of errors.

### 2.1. Statistical errors

There exist robust, though slow, algorithms for generating independent gauge configurations. The typical sample size has been at best $\sim 200$ independent configurations. The quality of the signal depends very much on the observable, however for the best case of spectrum calculations this sample size is adequate to reduce errors to less than 10 percent.

### 2.2. Finite box size errors

The energy $E$ of a state in a finite box with periodic boundary conditions is shifted due to interactions with mirror sources. Lüscher has shown [5] that for large enough $L$ the corrections are exponentially damped as $\exp -cEL$ where $c \approx 1$ is a constant that depends on the state, but the onset of the exponential regime has to be determined numerically. Present calculations indicate that for $E_{min}L \geq 4$ the asymptotic relation applies and that the errors are roughly a few percent.

## 2.3. Finite lattice spacing errors

The continuum action is the first term in a Taylor series expansion of the lattice action. At the classical level corrections start at $O(a)$ for the Wilson formulation of the Dirac term and $O(a^2)$ for staggered fermions. They are $O(a^2)$ for the gauge part. In addition there are $O(a)$ corrections in the operators used to probe the physics. These corrections can be large on accessible lattices (typically $a$ is in the range of $0.1 - 0.05$ fermi). There is considerable effort being made in the lattice community to reduce these errors by improving the lattice action and operators. It turns out that matrix element calculations are most severely affected by these $O(a)$ artifacts which are at present the largest source of uncertainty. In spectrum measurements these errors are much smaller once $a < 0.1$ fermi.

## 2.4. Extrapolations from heavier quarks

The quark propagator is the inverse of the Dirac operator. In the limit $m_q \to 0$ iterative algorithms used to calculate the inverse face critical slowing down. Since physical $u$ and $d$ quark masses are very nearly zero, and because over 90% of the time in QCD simulations is spent in calculating the inverse one has had to resort to extrapolating to the physical point from heavier masses (typically from $O(m_s)$ to $(m_u + m_d)/2 \approx m_s/25$). The functional form used in the extrapolation is usually derived using just the lowest order chiral perturbation theory. This procedure introduces systematic errors.

## 2.5. Effects of dynamical fermions

Simulations with dynamical fermions are prohibitively slow. As a result one works with the quenched approximation. This is a priori a totally uncontrolled approximation and I discuss it in more detail in the next Section.

## 2.6. Relation between lattice and continuum operators

In order to compare lattice results with those in the continuum we have to determine the relative normalization of the lattice and continuum operators. This is usually done using 1-loop perturbation theory, which leaves open the possibility that the 2-loop effects are large or there are large non-perturbative effects. A recent analysis by Lepage and Mackenzie suggests that 1-loop perturbation theory works very well provided one uses an appropriate definition of the coupling constant and one takes care of unwanted ultraviolet fluctuations using mean-field improvement [6]. So far the results from this approach agree very well with non-perturbative estimates in cases where the latter calculations are feasible. Further checks are under way.

## 3. Quenched versus unquenched calculations

In lattice QCD one calculates physical quantities as a statistical average over a set of background gauge configurations. For any given observable $\mathcal{O}$,

$$\langle \mathcal{O} \rangle = \frac{1}{Z} \int \Pi_{i,\mu} dU_{i,\mu} \, \mathcal{O}[U] \, \det M[U] \, e^{-S_g} \tag{3.1}$$

where $U_{i,\mu}$ is an SU(3) matrix defining the gauge field on a link in direction $\mu$ at site $i$. The background gauge configuration, $\{U_{i,\mu}\}$, is generated with Boltzmann weight $\det M[U] \, e^{-S_g}$. The factor $\det M[U]$ is the determinant of the Dirac operator and arises as a result of integrating over the quark degrees of freedom. Physically this factor takes into account the possibility that the QCD vacuum can create and annihilate quark/anti-quark pairs spontaneously. The determinant is a completely non-local object even though the initial Dirac action is only nearest-neighbor, and computationally very hard to include in the Monte Carlo procedure. It is therefore expedient to make an approximation – called the quenched approximation – in which one sets $\det M[U] = 1$. This corresponds to altering the QCD vacuum by artificially turning off vacuum polarization effects. The question to address then is how serious is this approximation.

The quenched vacuum possesses all three unique properties of QCD, *i.e.* confinement, asymptotic freedom and spontaneous chiral symmetry breaking. For this and other reasons it is expected that setting $\det M[U] = 1$ is a good approximation (on the level of 10%) for a large number of observables. Present simulations bear out this belief for sea quark masses roughly $\geq m_s$. While this is encouraging, it is by itself not sufficient to validate the approximation as sea quark effects in the same quantities are expected to be significant only for $m_q < m_s$. For this reason one has to proceed case by case, and eventually check using the full theory.

These checks are made difficult by the presence of statistical and systematic errors ( like finite lattice size and spacing, and extrapolation from heavier quarks) discussed above. Therefore, to expose the effects of vacuum polarization one needs to first bring these other errors down to the level of a few percent. Since the methodology for measuring many quantities is identical with or without the use of the quenched approximation to produce the statistical sample of background configurations, the strategy has been to first understand and control these errors in the simpler case. Thus the quenched approximation should be regarded as a test of our numerical techniques as well as a very good approximation to systematically improve upon.

The quenched approximation does have its limitations. Recent analysis, using chiral perturbation theory, of proton and pion masses show that in the quenched approximation these quantities develop non-analytic terms in addition to the desired physical behavior [7] [8]. So far it has been hard to exhibit the presence of these unwanted terms in numerical data; the hope is that the coefficients of these terms become significant only at much smaller quark masses and extrapolations from heavier masses are still sensible. Clearly this aspect of the quenched approximation needs more attention.

Let me end this discussion with a rough comparison of simulation time with and without dynamical fermions. With present algorithms the CPU requirements increase as $L^6$ for the quenched approximation and as $L^{10.5}$ with light dynamical fermions. Folding in the prefactors we find that for two degenerate flavors of quarks with roughly the mass of the strange quark, full QCD simulations are a factor of $1000 - 2000$ times slower. For smaller quark masses this factor will increase according to the above scaling behavior. As a result it is clear that we need improvements in update algorithms before contemplating realistic simulations with the full theory for the purpose of evaluating matrix elements within states made up of light hadrons.

## 4. Lattice QCD is not an open-ended problem

The masses of hadrons are very well measured experimentally. For this reason we know the different energy scales in the problem. To analyze the physics of light quarks ($u$, $d$, $s$) there are three scales that we have to consider. First $L > \xi_{maximum}$, and we take $\xi_{maximum} = 1/m_\pi$ as the pion is the lightest particle. Current simulations tell us that for $L/\xi_{maximum} \sim 5$ the finite size effects are down to a few percent level. Second, the lattice should be fine enough such that no essential features of the hadron's structure are missed as a result of discretizing the theory. This scale is controlled by $\xi_{minimum}/a$. We choose $\xi_{minimum}$ to be the reciprocal of the proton mass. Again current numerical data tell us that for $\xi_{minimum}/a \sim 5$ finite lattice spacing errors are reduced to the level of a few percent. Lastly, $\xi_{maximum}/\xi_{minimum} = M_{proton}/M_\pi = 7$ is an accurately measured number (getting this ratio correct in lattice simulations is equivalent to tuning $m_u$ to its physical value). Putting these three factors together tells us that definite measurements require lattices of size $L \sim 175$. Thus, unless present analysis has lead us to grossly underestimate the first two scales, definite calculations can be done in the quenched approximation on computers that can sustain 1-10 teraflops.

## 5. Hadron Spectrum

The first step towards the analysis of matrix elements is to calculate quark propagators. These quark propagators are combined to form hadron correlators. Matrix elements are calculated by sandwiching the appropriate operator between the initial and final state hadrons. The quality of the results depends on how well one has isolated the desired hadronic states before inserting the operator, for example eliminated the radial excitations that contaminate the signal. To extract the matrix element from the correlation function one has to remove the external legs by dividing the 3-point function by 2-point functions. Thus, a necessary condition for getting accurate results is to enhance the signal in the 2-point correlators –– quantities from which we extract decay constants and the energy of the state. It is therefore appropriate that as a prelude to presenting results for matrix elements I give a brief review of spectrum calculations.

Calculations of the light hadron spectrum use three input parameters; two quark masses, $m_u$ and $m_s$ (we assume $m_u = m_d$), and the bare gauge coupling constant. The quark masses are adjusted to give the physical masses for the $\pi$ and $K$ mesons. In practice one adjusts the ratio of their mass to that of the proton and, as mentioned above, at present we have to make an extrapolation from heavier quark masses. If QCD is the correct theory of strong interactions then all other mass ratios should agree with experimental numbers as the bare gauge coupling is tuned to zero. Again we extrapolate $g_{bare} \to 0$ using renormalization group scaling. The status of these calculations is summarized by Ukawa at LATTICE92 meeting [9], and the most complete calculation to date is by Butler *et al.* [10].

The results show that finite size errors are down to a few percent level when $L/\xi_{maximum} \geq 5$ and finite lattice spacing errors are of similar size for $\xi_{minimum} \geq 5$. More importantly, the quenched results agree with experimental data to within 10%. This is a remarkable agreement considering the shift in rho mass due to $\rho \to \pi\pi$ decay has not been taken into account in setting the scale. For this reason I would like to see independent confirmation of the results of Butler *et al.* before declaring this aspect of spectrum calculations under control. In any case these results, in part, form the basis of my earlier conclusions on relevant scales. The finite $a$ errors are expected to be much larger in matrix element calculations as discussed later.

## 6. Semi-leptonic form factors of heavy-light mesons from lattice QCD

The semi-leptonic decays of mesons containing one heavy valence quark ($c$, $b$) and one light valence quark ($u$, $d$, $s$) may provide the most accurate determination of the flavor mixing angles. Consider the case, $D \to X l \nu$, where $X$ has flavor content $\overline{u}s$ ($K$ or $K^*$). In the one $W$ exchange approximation the amplitude is

$$\begin{aligned} \langle X^- l^+ \nu | H_W | D^0 \rangle &= \frac{G_F}{\sqrt{2}} \int d^4x \, \langle X^- l^+ \nu | (V-A)^\dagger_\mu (V-A)_\mu | D^0 \rangle, \\ &= \frac{G_F}{\sqrt{2}} V_{sc} \overline{v}(l) \gamma_\mu (1-\gamma_5) u(\nu) \langle X^- | \overline{s} \gamma_\mu (1-\gamma_5) c | D^0 \rangle, \end{aligned} \quad (6.1)$$

where $G_F$ is the Fermi constant, $V_{cs}$ is the $c \to s$ CKM matrix element. This process is particularly simple because the hadronic and leptonic currents factorize. The leptonic part of the decay can be calculated accurately using perturbation theory, while to take into account non-perturbative contributions to the hadronic part

$$H_\mu = \langle X|\bar{s}\gamma_\mu(1-\gamma_5)c|D\rangle \qquad (6.2)$$

one resorts to lattice QCD. In this talk I will present our results for the case $D^0 \to K^-e^+\nu$ as it is the simplest.

6.1. $D^0 \to K^-e^+\nu$

The matrix element $H_\mu$ can be parameterized in terms of two form factors:

$$\langle K^-(p_K)|\bar{s}\gamma_\mu(1-\gamma_5)c|D^0(p_D)\rangle = p_\mu f_+(Q^2) + q_\mu f_-(Q^2), \qquad (6.3)$$

where $p = (p_D + p_K)$ and $q = (p_D - p_K)$ is the momentum carried away by the leptons, and $Q^2 = -q^2$ (which is always positive). I use the Euclidean notation $p = (\vec{p}, iE)$ so that $p^2 = \vec{p}^2 - E^2$. An alternative parameterization is

$$\langle K^-(p_K)|\bar{s}\gamma_\mu(1-\gamma_5)c|D^0(p_D)\rangle$$
$$= \left(p_\mu - \frac{m_D^2 - m_K^2}{Q^2}q_\mu\right)f_+(Q^2) + \frac{m_D^2 - m_K^2}{Q^2}q_\mu f_0(Q^2), \qquad (6.4)$$

where

$$f_0(Q^2) = f_+(Q^2) + \frac{Q^2}{m_D^2 - m_K^2}f_-(Q^2). \qquad (6.5)$$

In the center of mass coordinate system for the lepton pair, i.e. $\vec{q} = 0$ or equivalently $\vec{p}_K = \vec{p}_D$, one has

$$\langle K^-(p_K)|\bar{s}\vec{\gamma}c|D^0(p_D)\rangle = 2\vec{p}_D f_+(Q^2),$$
$$\langle K^-(p_K)|\bar{s}\gamma_4 c|D^0(p_D)\rangle = \frac{m_D^2 - m_K^2}{\sqrt{Q^2}}f_0(Q^2). \qquad (6.6)$$

Thus, the form factor $f_+(Q^2)$ is associated with the exchange of a vector particle, while $f_0(Q^2)$ is associated with a scalar exchange. It is common to assume nearest pole dominance and make the hypothesis

$$f_+(Q^2) = \frac{f_+(0)}{1 - Q^2/m_{1^-}^2}, \qquad f_0(Q^2) = \frac{f_0(0)}{1 - Q^2/m_{0^+}^2}, \qquad (6.7)$$

where $m_{J^P}$ is the mass of the lightest resonance with the right quantum numbers to mediate the transition; $D_s^+(1969)$ or $D_s^{*+}(2110)$ in the pseudoscalar or vector channels respectively. The goal of the lattice calculations is to determine the normalizations $f_+(0)$ and $f_0(0)$ and map out the $Q^2$ dependence.

In the limit of vanishing lepton masses, the vector channel dominates and one can write the the differential decay rate as

$$d\Gamma(Q^2) = \frac{G_F^2 |V_{cs}|^2}{192\pi^3 m_D^3} dQ^2 \lambda(Q^2)^{3/2} |f_+(Q^2)|^2,$$
$$\lambda(Q^2) = (m_D^2 + m_K^2 - Q^2)^2 - 4m_D^2 m_K^2.$$
(6.8)

To integrate this, the functional form of $f_+$ must be known. Assuming vector meson dominance numerical integration gives

$$\Gamma(D^0 \to K^- e^+ \nu) = 1.53 |V_{cs}|^2 |f_+(0)|^2 \times 10^{-11} \text{sec}^{-1}.$$
(6.9)

Eqn. (6.9) is the simplest example of the master equation; using it we can extract $V_{cs}$ once $\Gamma(D^0 \to K^- e^+ \nu)$ has been measured and $f_+$ calculated using lattice QCD. In this case, however, $|V_{cs}| = 0.975$ is known very accurately, so one extracts $|f_+(0)| \approx 0.75$. The quantity $f_0(0)$ has not been determined.

The details of our lattice calculation of the form-factors are given in Ref. [11], so here I briefly describe some of the lattice technicalities and present the results. I would like to emphasize that the results presented here are exploratory. The goal was to investigate different numerical techniques in order to improve the signal to noise ratio. The data confirm that the numerical techniques are now good enough to get reliable results with today's massively parallel computers.

### 6.2. Lattice parameters

Our statistical sample consists of 35 lattices of size $16^3 \times 40$ at $\beta = 6.0$ corresponding to a lattice spacing $a = 0.1$ fermi. We fix the heavy (charm) quark mass at $\kappa = 0.135$, and use only two values of the light quark mass, $\kappa = 0.154$ and $0.155$. Using $a^{-1} \cong 1.9 \ GeV$, this corresponds to a heavy-light meson of mass 1.59 and 1.54 GeV (about the mass of the physical charm quark) and to light-light pseudoscalar masses of roughly 690 MeV and 560 MeV. Our heavy-light pseudoscalar mesons therefore correspond most closely to the physical $D$ meson, with a somewhat massive light constituent, while the light-light mesons are analogous to the physical $K$. We will henceforth adopt this nomenclature.

### 6.3. Quark propagators and 3-point Correlation function

The calculation of quark propagators is done on lattices doubled in the time direction, i.e. $16^3 \times 40 \to 16^3 \times 80$. We use periodic boundary conditions in all four directions. These propagators on doubled lattices are identical to forward and backward moving solutions on the original $16^3 \times 40$ lattice. To improve the signal we use the "Wuppertal" smeared source method for generating the propagators.

In the 3-point correlation function the source for the $K$ meson is fixed at $t_K = 1$ and for the $D$ meson at $t_D = 32$. As a result the wrap-around effects in time direction are exponentially damped by at least 18 time slices because of doubling the lattices. The position of the insertion of the vector current is varied over $4 < t < 28$ to improve the statistics. The lowest order Feynman diagram for this process is shown in Fig. 1a. Fig. 1b shows one possible correction term due to gluon interactions which make perturbative analysis of the matrix element hard.

*6.4. Operators and correlators*

In order to get a handle on $O(a)$ effects coming from the lattice operator we use three transcriptions for the vector current

$$V_\mu^{\text{local}}(x) = \overline{q}_1(x)\gamma_\mu q_2(x),$$
$$V_\mu^{\text{ext.}}(x) = \frac{1}{2}\left(\overline{q}_1(x)\gamma_\mu U_\mu(x) q_2(x+a\mu) + \overline{q}_1(x+a\mu)\gamma_\mu U_\mu(x)^\dagger q_2(x)\right), \quad (6.10)$$
$$V_\mu^{\text{cons.}}(x) = \frac{1}{2}\left(\overline{q}_1(x)(\gamma_\mu - 1)U_\mu(x)q_2(x+a\mu) + \overline{q}_1(x+a\mu)(\gamma_\mu + 1)U_\mu(x)^\dagger q_2(x)\right).$$

In our calculation the quarks $q_1$ and $q_2$ may both be light, or one heavy and one light. Note that $V_\mu^{\text{cons.}}(x)$ is conserved only for degenerate quarks. We use the Lepage-Mackenzie improved normalization of these currents relative to the continuum vector current. The lattice field for a quark of flavor $i$ is related to its continuum counterpart by

$$\psi_{cont}^i = \sqrt{8\kappa_c}\sqrt{1 - \frac{3\kappa_i}{4\kappa_c}}\psi_L^i \quad (6.11)$$

where $\kappa_c = 0.15702$ is the value of the hopping parameter that corresponds to zero pion mass. To get the normalization of the local vector current we multiply the 1-loop perturbative result for the operator by that for $8\kappa_c$. This gives a better perturbative expansion as the large tadpole contributions (lattice artifacts) are cancelled. The result is

$$V_\mu\bigg|_{cont} \equiv \overline{q}_1(x)\gamma_\mu q_2(x)\bigg|_{cont} = \sqrt{1 - \frac{3\kappa_1}{4\kappa_c}}\sqrt{1 - \frac{3\kappa_2}{4\kappa_c}}(1 - 0.82\alpha_V)\overline{q}_1(x)\gamma_\mu q_2(x)\bigg|_L, \quad (6.12)$$

where $\alpha_V = g_R^2/4\pi$ is the renormalized coupling, which we take to be $g_R^2 = 1.7 g_{bare}^2$.

In the extended 1-link and conserved currents the tadpoles cancel, and to $O(\alpha_s)$ the relation between continuum and lattice operators is (the details are given in Ref. [11])

$$V_\mu\bigg|_{cont} = 8\kappa_c\sqrt{1 - \frac{3\kappa_1}{4\kappa_c}}\sqrt{1 - \frac{3\kappa_2}{4\kappa_c}}(1 - 1.038\alpha_V)V_\mu^{ext.}\bigg|_L, \quad (6.13)$$

and similarly for the conserved current

$$V_\mu\bigg|_{cont} = 8\kappa_c\sqrt{1 - \frac{3\kappa_1}{4\kappa_c}}\sqrt{1 - \frac{3\kappa_2}{4\kappa_c}}V_\mu^{cons.}\bigg|_L. \quad (6.14)$$

In the next sub-section I present our data and demonstrate that to get consistent results between the three lattice currents it is important to use these normalizations.

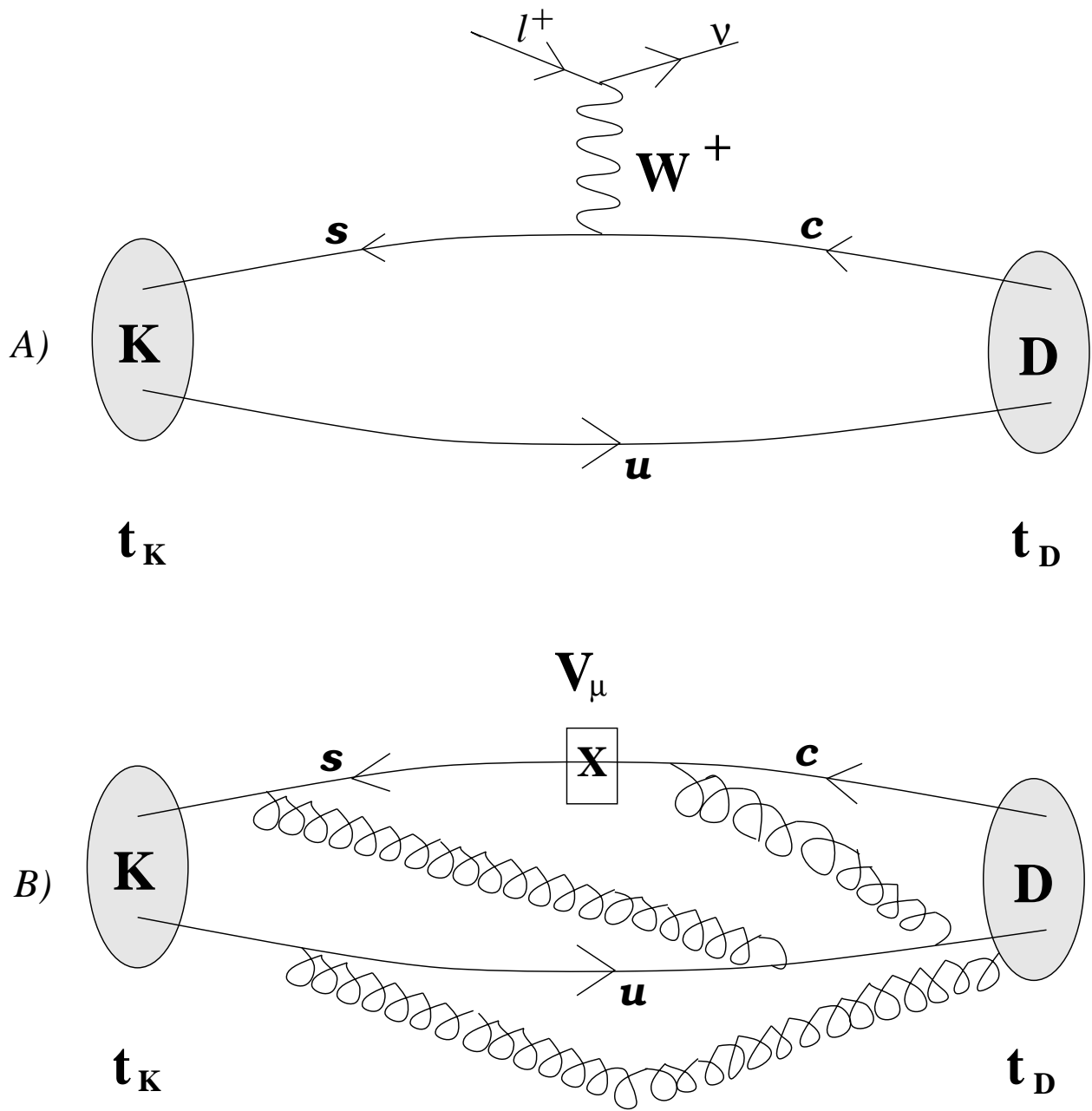

**Fig. 1.** (A) The semi-leptonic decay of a $D^0$ meson to a $K^-l^+\nu$ final state. The $c \to s$ transition takes place through the emission of a $W^+$ and only the vector part of the $V - A$ weak current contributes. The interaction is not pointlike at the hadronic vertex and its $q^2$ dependence is given by the form-factors. (B) An example of QCD corrections to the matrix element $H_\mu$.

*6.5. Results*

I am going to skip over all the details of the analysis and the discussion of the quality of the signal in the correlators due to lack of time. These are given in Ref. [11]. The final results for the form-factors are given in Table 1. Our analysis show that within their respective 1-$\sigma$ uncertainty the three different lattice transcriptions of the vector current give consistent results and the difference between the local, extended and "conserved" currents can be taken to be a measure of the remaining $O(a)$ corrections. The numbers do not show a large variation for the two values of the light quark mass that we have used and the value of $f_+(Q^2)$ is roughly consistent with the phenomenological value $f_+(0) = 0.75$.

| \multicolumn{5}{c}{$\kappa = 0.154$} | | | | |
|---|---|---|---|---|
| Current | $f_+(Q^2 = 0.217)$ | $f_-(Q^2 = 0.217)$ | $f_0(Q^2 = 0.217)$ | $f_0(Q^2 = -0.05)$ |
| $V_\mu^{Local}$ | 0.61(11) | $-0.44(25)$ | 0.66(13) | 0.91(9) |
| $V_\mu^{Ext.}$ | 0.68(12) | $-0.41(24)$ | 0.72(14) | 1.01(11) |
| $V_\mu^{Cons.}$ | 0.80(12) | $-0.30(23)$ | 0.83(13) | 1.18(12) |
| \multicolumn{5}{c}{$\kappa = 0.155$} | | | | |
| Current | $f_+(Q^2 = 0.260)$ | $f_-(Q^2 = 0.260)$ | $f_0(Q^2 = 0.260)$ | $f_0(Q^2 = -0.035)$ |
| $V_\mu^{Local}$ | 0.65(20) | $-0.65(36)$ | 0.69(21) | 0.96(10) |
| $V_\mu^{Ext.}$ | 0.66(24) | $-0.52(36)$ | 0.70(24) | 1.04(11) |
| $V_\mu^{Cons.}$ | 0.80(27) | $-0.37(38)$ | 0.82(27) | 1.23(13) |

Table 1: The data for semi-leptonic form-factors for each of the three definitions of the lattice vector current. The two values of light quark mass correspond to pions of roughly 690 and 560 MeV.

We can also compare our results with earlier calculations as these were done with similar lattice parameters. The group of Bernard *et al.* [12] measured the form-factors on $24^3 \times 40$ lattices at the same values of $\beta$ and $\kappa$. They used only the local vector current, and adopted a different normalization. Converting their result to the normalization we use gives $f_0(\vec{p} = 0) = 0.85(10)$ at $\kappa = 0.154$ to be compared with our value of 0.91(9). Similarly the Rome-Southampton group [13] [14] have measured the form-factors on $20 \times 10^2 \times 40$ lattices at the same value of $\beta$ and similar $\kappa$. They use the "conserved" 1 vector current. Again, using the same normalization for the vector current that we use and interpolating their results to $\kappa = 0.154$, we find $f_+(\vec{p} = 2\pi/L) = 0.72(7)$ to be compared with our result of 0.80(12) and $f_0(\vec{p} = 2\pi/L) = 0.70(5)$ to be compared with 0.83(13).

The internal consistency of our results and the agreement with previous calculations shows that semi-leptonic form-factors can be extracted from lattice simulations. The largest source of error in present results comes from $O(a)$ corrections and an inadequate signal in the non-zero momentum correlators. The next round of calculations are being done on $32^3 \times 64$ lattices on the CM5. These will hopefully address the phenomenologically interesting cases of the decay of $D$ to vector mesons and of $B \to \pi$ and $B \to D$ which are crucial for extracting $V_{bu}$ and $V_{bc}$ from the experimental data.

## 7. The kaon B parameter

CP violation in the standard model is governed by a single parameter $\delta$ provided we assume that $\Theta = 0$. Once the value of $\delta$ is known then each CP violating process will provide a constraint involving the mixing angles and quark masses. I illustrate this using as an example the mixing between $K^0$ and $\overline{K^0}$ as it is the best measured CP violating process.

The mass eigenstates in the neutral kaon system are defined as

$$\begin{aligned} |K_L\rangle &= \frac{1}{N} \left[ (1+\epsilon)|K^0> + (1-\epsilon)|\overline{K}^0> \right] \\ |K_S\rangle &= \frac{1}{N} \left[ (1+\epsilon)|K^0> - (1-\epsilon)|\overline{K}^0> \right] \end{aligned} \qquad (7.1)$$

where $N$ is the normalization. The parameter $\epsilon$ measures the amount of CP violation, and in the standard model is given by the master equation [2]

$$\epsilon = 1.4 e^{i\pi/4} \sin\delta B_K \left\{ [\eta_3 f_3(m_c, m_t) - \eta_1] \frac{m_c^2}{m_W^2} + \eta_2 \frac{m_t^2}{m_W^2} f_2(m_t) \mathrm{Re}\left(\frac{V_{td}^* V_{ts} V_{ud} V_{us}^*}{s_{12}^2}\right) \right\} \quad (7.2)$$

where $\eta_1 = 0.7$, $\eta_2 = 0.6$ and $\eta_3 = 0.4$ are the QCD correction factors and $f_2$ and $f_3$ are known functions of the quark masses. The value of $\epsilon$ is known experimentally to be

$$|\epsilon| = (2.258 \pm 0.018) \times 10^{-3}. \qquad (7.3)$$

In Eq. (7.2) the strong interaction corrections are encapsulated in the parameter $B_K$ which is the ratio of the matrix element of the $\Delta S = 2$ four-fermion operator $(\overline{s}\gamma_\mu(1-\gamma_5)d)(\overline{s}\gamma_\mu(1-\gamma_5)d)$ to its value in the vacuum saturation approximation

$$\langle \overline{K^0} | (\overline{s}\gamma_\mu(1-\gamma_5)d) | 0 \rangle \langle 0 | (\overline{s}\gamma_\mu(1-\gamma_5)d) | \overline{K^0} \rangle = \frac{16}{3} f_K^2 M_K^2 B_K . \qquad (7.4)$$

Theoretical estimates of this parameter vary from 0.33 to 1 and lattice calculations aim to provide a non-perturbative answer.

The steps in the calculation leading to Eqn. (7.2) are show in Fig. 2. In the standard model $K^0\overline{K^0}$ mixing can occur due to the second order weak process shown in Fig. 2a. Since the $W^\pm$ and the top quark are heavy, it is expedient to integrate them out and define an effective 4-fermion interaction at some scale $\mu > m_c$. This is represented by the diagram in Fig. 2b. This weak amplitude is modified by strong interaction corrections as illustrated in Fig. 2c, and it is these corrections that change the value of $B_K$ from 1.0.

The calculation of $B_K$ has been done with both staggered and Wilson fermions. At present simulations using staggered fermions are far more extensive and have much less theoretical uncertainty. The two formulations give consistent results [15], so I will present results only for staggered fermions as these have much smaller errors. The details of these calculations are given in Refs. [16] [17] [18]. Our final results from different lattices and for different values of $a$ are shown in Fig. 3. This calculation is sufficiently mature that one can analyze the data with respect to the 6 sources of errors discussed in Section 2.

1. *Statistical errors:* Three independent samples of configurations have been analyzed at $\beta = 6.0$ and results for $B_K$ are consistent within errors. Also, the Japanese group [18] have carried out a totally independent calculation and get the same results. I take this to indicate that the analysis of statistical errors is correct.

2. *Finite Size errors:* We have compared results on $16^3 \times 40$ lattices with those on $24^3 \times 40$ at $\beta = 6.0$ and on $18^3 \times 42$ lattices with those on $32^3 \times 48$ at $\beta = 6.2$. In both cases the results are consistent. Our conclusion is that finite size effects in the data presented in Fig. 3 are much smaller than the statistical errors and at most $1-2\%$.

3. *Finite lattice spacing errors:* These errors come from both the lattice action and the operators used in the measurements. Fig. 3 shows two different extrapolations assuming corrections to be either $O(a)$ or $O(a^2)$. These two different ways of extrapolation yield $B_K g^{-4/9} = 0.44(4)$ versus $0.54(2)$ in the continuum limit. The uncertainty in the form of extrapolation to use is at present the largest source of error in the data. Preliminary analysis suggests that the corrections in staggered fermion data are $O(a^2)$. This will be checked by improving the statistics at $\beta = 6.4$ and doing another simulation at, say, $\beta = 6.6$.

4. *Extrapolation in $m_q$:* The $K^0$ consists of $d$ and $\overline{s}$ valence quarks. In our calculations the values of $B_K$ are read off from a simulation in which the two quarks are almost degenerate, say both with mass $m_s/2$. We have done some tests by varying the two quark masses in the range $m_s/3 - 3m_s$ to check for effects of using non-degenerate masses. So far our conclusion is that these are at best a few percent. Going to smaller masses becomes increasingly harder as it requires higher statistics and a larger lattice, but otherwise the calculation is the same.

5. *Quenched approximation:* Two independent calculations have been done using

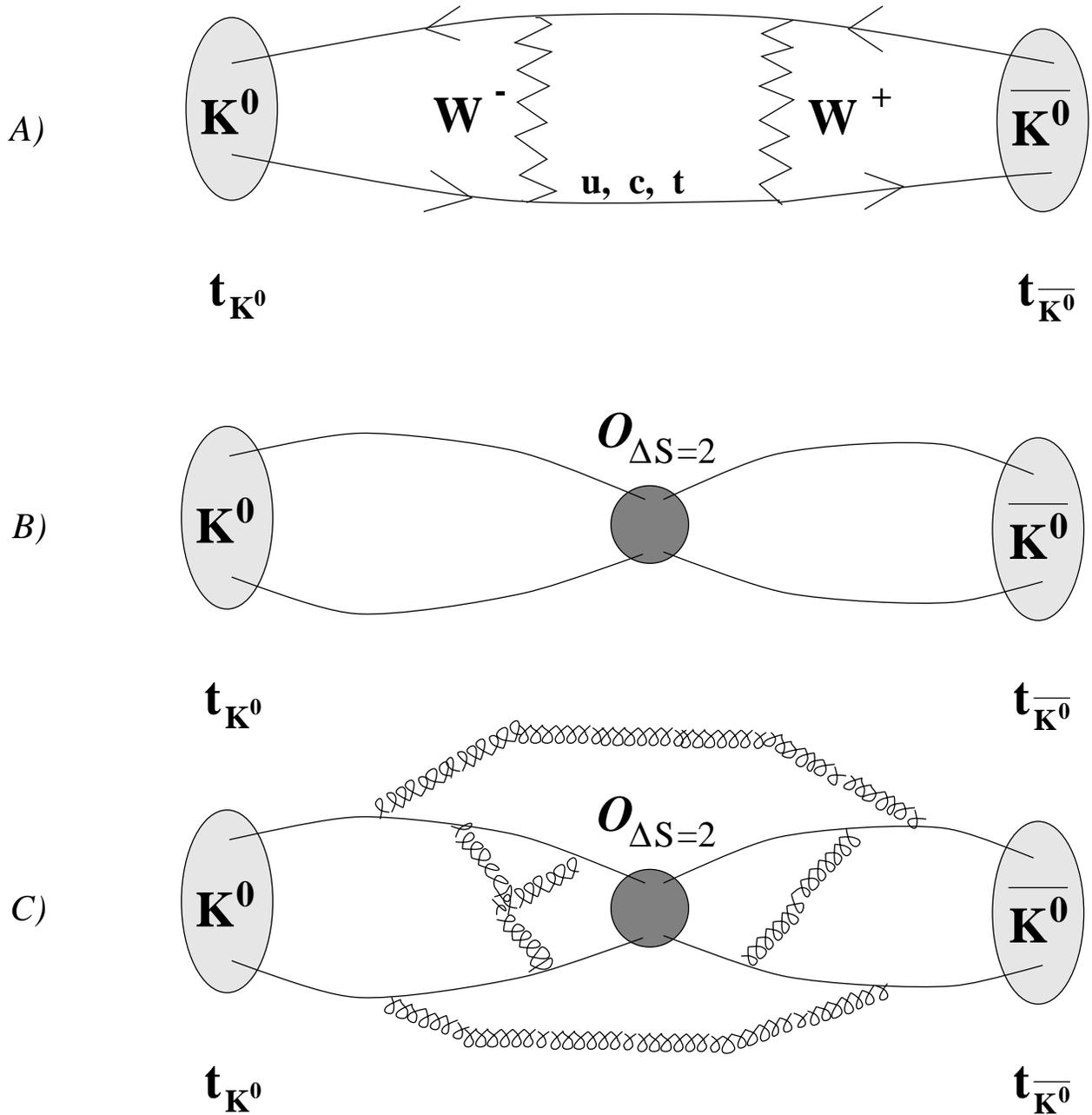

**Fig. 2.** (A) One of the two possible box diagrams responsible for the mixing between $K^0\overline{K^0}$. (B) The short distance interactions involving the $W$ exchange and $t$ quark intermediate state is replaced by the $\Delta S = 2$ 4-fermion effective interaction. (C) One possible QCD correction to the weak decay. Lattice QCD is a non-perturbative method to sum all such possible corrections.

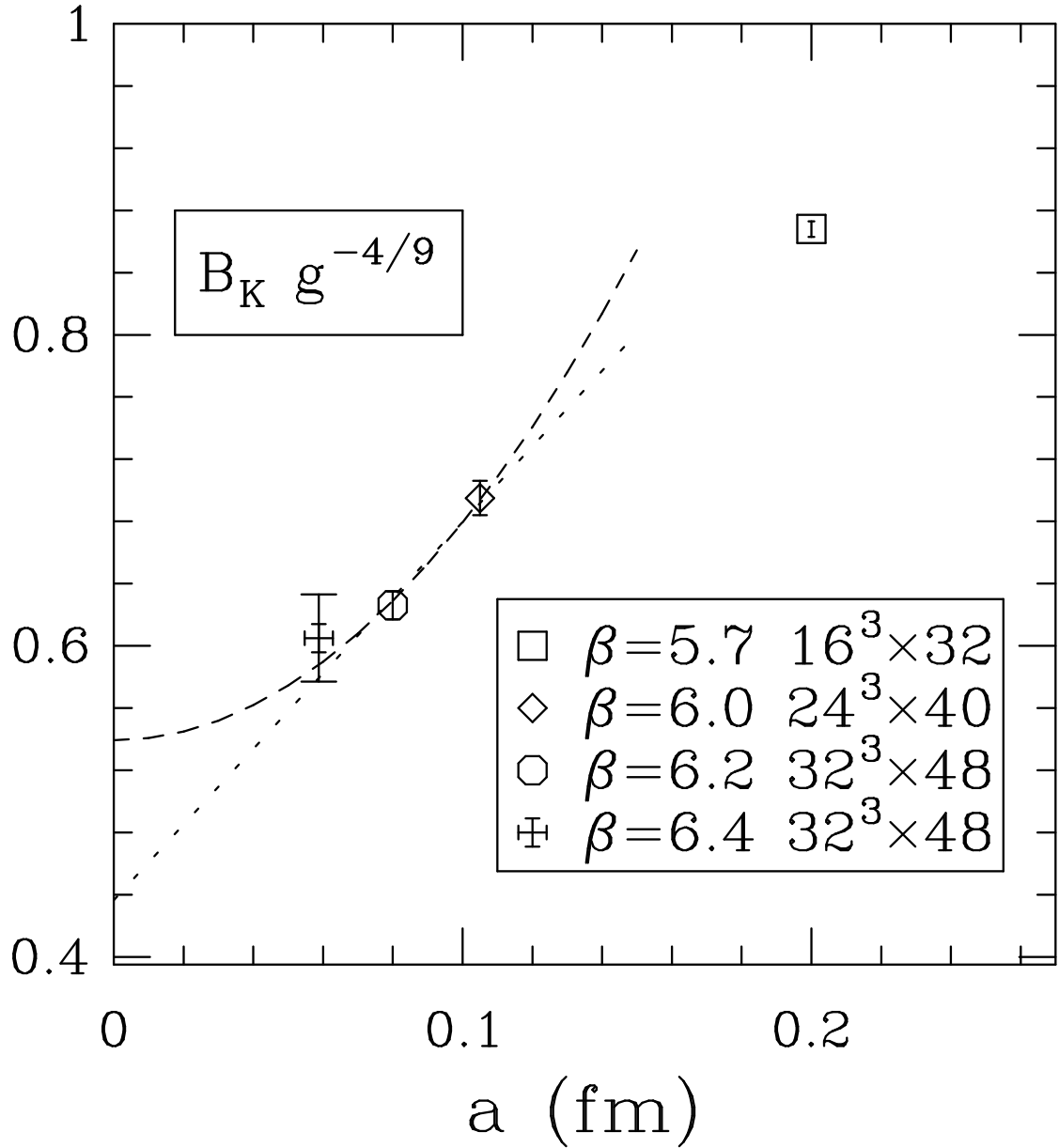

**Fig. 3.** The result for $B_K$ as a function of the lattice spacing $a$. The factor $g^{-4/9}$ is the relative scaling factor for the different values of $\beta$.

lattices generated with 2 flavors of dynamical fermions [19] [20]. The quark mass in the update is $\sim m_s$. The results, though preliminary, are consistent within errors with the quenched data. Based on this comparison our present estimate is that quenching may introduce only a $5-10\%$ correction, making $B_K$ one of the first quantities for which we expect lattice QCD to yield accurate results. To improve upon this first check we need to study the effect of tuning $m_d$ to its physical value both in the update of lattices and in the valence quark propagator.

6. *Operator renormalization:* The 1-loop calculation relating the lattice operator to the continuum has been done [21] [18], and the upshot of it is that including this factor reduces $B_K$ by about $6-7\%$.

Finally, to make contact with phenomenology we have to remove the dependence on the renormalization point $\mu$ at which the effective theory is defined in the continuum. The $\mu$ independent parameter is $\widehat{B}_K = B_K \alpha_s^{-2/9}$, and for $\beta = 6.0$ the correction factor is $\alpha_s^{-2/9} = 1.34$ with roughly a 10% uncertainty coming from the uncertainty in the lattice scale [22].

With all these estimates in hand our current estimate is $\widehat{B}_K = 0.68(10)$. To get this I have used the $O(a^2)$ extrapolation for $B_K$ data and have only included the operator renormalization factor as the other sources of systematic errors are smaller and less well determined.

To conclude, I hope I have convinced you that lattice QCD calculations can play a very important role in our understanding of the standard model. The quality of results will be systematically improved with better numerical techniques and with bigger and faster computers. Therefore it is appropriate that I end this talk with a brief report on the status and performance of our QCD codes on the CM5.

## 8. Optimization of QCD codes on the CM5

We have finished the first phase of the development of QCD codes on the CM5. The overall strategy is to keep all the control structure in CMFortran under the SIMD programming environment. We isolate the computationally intensive portions of the code and convert them to CDPEAC. This way we are able to preserve modularity in order to implement changes in the algorithm and to add new measurement routines very quickly.

The two key operations that capture the essence of QCD calculations are

$$\begin{aligned} A &= B + C * D \\ A &= B + C * cshift(D) \end{aligned} \qquad (8.1)$$

where $A$, $B$, $C$, $D$ are $3 \times 3$ complex matrices and the circular shift (cshift) is by $\pm 1$ lattice units in one of the four directions. (Same amount of communication is done in all four directions). The lattice size being used is $32^3 \times 64$ and we use single precision variables. Thus a typical array layout is $A(:serial,:serial,:news,:news,:news,:news)$ with dimensions $A(3, 3, 32, 32, 32, 64)$. At present the second operation is broken up into two parts

$$\begin{aligned} tmp &= cshift(D) \\ A &= B + C * tmp \end{aligned} \quad (8.2)$$

as there is no way to overlap communications with computations at the CMF level. The key lessons learned from optimizing the above two kinds of primitives are*:

1. There is no discernible performance penalty for calls to CDPEAC routines. So the code can be made modular and portable by converting small compute intensive parts into CDPEAC subroutines.

2. We vectorize over the sites. All loads and stores are joined with arithmetic operations, so we reload variables as necessary. This allows us to optimize register use to get a long vector length.

3. Each time we load a different array, say $B$ after $C$, we pay a penalty of 5 cycles due to DRAM page faults. Since data elements in a vector load are contiguous in memory, there is no penalty within the vector operation. The DRAM page faults reduce the maximum possible speed from 64 to 50 MIPS/node. Other forms of data layout do not provide any significant improvement in performance and we do not recommend hand tuned layouts as they make the code much more complicated without any gain in speed.

4. For on node calculations we sustain approximately 50 Megaflops/node for multiplies or adds and 100 when we can chain multiply with add. Thus we are able to get optimal performance with very simple vectorization and data layout strategy.

5. By writing matrix multiply in CDPEAC we avoid single-precision loads and stores (this constitutes the bulk of the factor of $3-5$ performance gain over CM Fortran) as complex numbers are double word aligned. Single stores should be avoided whenever possible.

6. The cshift operation is slow due to off-node communication speed and because it does unnecessary memory to memory transfer of on-chip data. In SIMD mode the unnecessary moves can be avoided only by combining cshift with the matrix multiply. Also, part of the on-VU arithmetic can be done while the off-node data is in the network. This optimization step requires writing what is essentially a stencil in DPEAC, and we are currently implementing this with help from staff at Thinking Machines.

---

\* All tests and comparison timings were done using CM Fortran Driver Version: 2.1 Beta 1 Rev: f2100 w/ release 2.1 beta 0.1. A number of inefficiencies have been fixed in Version: 2.1 Beta 1 Rev: f2100 but we have not yet timed our codes under it.

In conclusion, it is clear that to develop an optimizing CMF compiler is hard and performance aficionados will have to program at CDPEAC level for possibly the complete lifetime of the present architecture. Therefore, I have not discussed any of the inefficiencies of CMF that are removed by writing in CDPEAC. For those who are willing to write in CDPEAC there is additional reward as the CM5 is a stable high performance massively parallel computer.

## Acknowledgements

It is a pleasure to acknowledge my collaborators Tanmoy Bhattacharya, David Daniel, Greg Kilcup, Apoorva Patel and Steve Sharpe in the calculations presented here. The development of QCD codes on the CM5 has been spearheaded by Tanmoy Bhattacharya, and without him we would have made little progress in CDPEAC optimization. We acknowledge the tremendous support received from the DoE HPCC initiative and a Grand Challenge award at the Advanced Computing Laboratory at Los Alamos. The National Center for Supercomputer Applications at Urbana-Champaign provided valuable resources in the development of CM5 codes and we thank L. Smarr for this opportunity. The results presented in this talk were obtained on Crays at NERSC, Los Alamos, PSC and SDSC, and we thank J. Mandula, R. Roskies and A. White for their enthusiastic support over many years. Finally, many thanks to Priya and Rajiv for showing me a wild side to the otherwise sleepy southern town.